# *Quantitative Paradigm of Software Reliability as Content Relevance*

Yuri Arkhipkin
Moscow, Russia
aryur@yandex.ru

This paper presents a quantitative approach to software reliability and content relevance definitions validated by the systems' potential reliability law. Thus it is argued for the unified math nature or quantitative paradigm of software reliability and content relevance.

This paradigm **integrates** subject matter data and quantitative software reliability metrics and is viewed as a foundation of the software engineering that provides continuous reliability evaluations throughout software development process, thus accounting and tracing quantitative software reliability requirements from customer to product.

This paradigm **integrates** subject matter data and quantitative content relevance metrics and is viewed as a foundation of the content relevance engineering that provides continuous relevance evaluations throughout content engineering process, thus, making possible to trace content relevance requirements from customer to product.

This **integrated** development environment is to be assured and supported by quantification grammar as a foundation of software reliability quantification programming language and content relevance quantification markup language.

## 1. Introduction

Much research was done on models approaching software reliability quantification. The results seem to be of poor satisfaction despite of the increasing number of these models. The lack of explicit evaluations of software elements' failure probability may be considered as one of the main problems in software reliability quantification. Software elements' failure data is yielded while testing (executing) software for a vast field of subject applications. This data may be considered as an ad hoc data much depending on developer's skill and software testing skill in particular. Software testing in general may be considered as a trial failure process of sensitizing software elements (sites) to define whether yielded results are true or faulty.

Much research was also done on models approaching content relevance quantification. The results seem to be of not enough satisfaction despite of the increasing number of these models. Content's grammar variety may be considered as one of the main problems in content relevance quantification. Content irrelevance (failure) data is yielded due to query occurrence (generation) through content searching, thus providing data for query's terms refinements and (or) search engine's improvements. So this data generally may be considered as an ad hoc data much depending on search engine's developer skill and query generation (testing) skill in particular. Content searching in general may be considered as a trial failure process of sensitizing content terms to define whether this content is relevant to the terms of the query or not.

This paper offers to break through the quantification problems of software reliability and content relevance engineering by approaching any digital content as a trial failure system regardless of its grammar and subject matter applications.

**Chapter 2** introduces briefly some mathematics of the systems' potential reliability law proved by B. S. Fleishman [1]. This law validates the systems' failure intensity quantitative ranges depending on known potential operating elements' number as a part of system elements' total number and their mean operating probability.

**Chapter 3** presents the quantitative approach to the software reliability engineering validated by systems' potential reliability law. Software site's operating probability is considered to be equal to the site's probability occurrence or potential occurrence. This probability may be evaluated at any software development cycle. Developer in general needs no external statistic data to monitor the achieved quantitative reliability level of the software project.

**Chapter 4** presents the quantitative approach to the content relevance engineering validated by systems' potential reliability law. Content element's operating (sensitizing sense) probability is considered to be equal to the term's content frequency. This frequency may be evaluated at any cycle of the content development. Developer in general needs no external statistic data to index the quantitative relevance level of the content project.

**Chapter 5** concludes with preliminary approach to software reliability quantification grammar to be implemented as a language of the integrated development environment for a product lifecycle solution in software engineering.

The pragmatics of the presented paradigm may be defined by its validity and verifiability that may be explicitly quantified for the vast field of system engineering applications.

## 2. Potential Reliability of a Trial Failure System

At every time moment the system's elements belong to either operating or failure state. Moreover the conversion occurs instantly from operating to failure state, while the reverse conversions are impossible.

It is natural in general to consider a system as an operating one at the given moment, if there exist at least some operating elements comprising a before stated minimal part of total system elements' number. Many uncontrollable causes, influencing elements' failures, make it possible to consider the failures' occurrence as random events.

Let the system $A_R$ at given moment $\tau$ consists of $n$ elements $a_1, \ldots, a_v, \ldots, a_n$ with arbitrary interactions. Any element is associated with two mutually exclusive events $A_1^v$ and $A_0^v$. The event $A_1^v$ is associated with operating element $a_v$ and the event $A_0^v$ is associated with its failure. Let the probabilities of the events $A_1^v$ and $A_0^v$ are equal to $p_v$ and $1-p_v$ correspondingly.

Consider the set $R_n$ of all possible $2^n$ states $\rho=(i_1, \ldots, i_v, \ldots, i_n)$ of the system $A_R$. This set depicts the operating and failure states of the system $A_R$ so that $i_v=1$ if $a_v$ is in the state $A_1^v$ and $i_v=0$ if $a_v$ is in the state $A_0^v$.

Let us divide the set $R_n$ into two parts $E_1$ and $E_0 = R_n \setminus E_1$. The set $E_1$ is an operating set of the system $A_R$ and $E_0$ is a failure set of the system $A_R$. Consider by the definition that the system $A_R$ operates at the given moment only if $\rho \in E_1$.

It is considered that the system's state $\rho$ is a sequence of independent trials with outcomes probabilities $p_v = P(i_v=1), 1-p_v = P(i_v=0) \ (v=1,2,\ldots,n)$ of every $v-th$ trial. Then the probability $Pv$ of the system $A_R$ to operate at the given moment may be defined [1] as:

$$P_v = P(\rho \in E_1) = \sum_{\rho \in E_1} \prod_{v=1}^{n} p_v^{i_v} \cdot (1-p_v)^{1-i_v} \qquad (2.1)$$

Consider the systems comprised of $n$ elements, operating set $E_1$ of which consists of states, each including more than $s$ operating elements. So the set $E_1$ includes all system's states $\rho=(i_1,\ldots,i_v,\ldots,i_n)$ for which $\sum_{v=1}^{n} i_v > s$. Such systems are named as symmetric of $s-th$ degree.

To define the possibility to operate for the symmetric system of $s-th$ degree, it is necessary to study the asymptotic behavior of (2.1) at $n \to \infty$.

Restricting the study by operational systems with large but constant elements' number $n$, it is said that these systems have instant operating probability $P_v(\tau)$ at the given moment $\tau$. The probability $R(t)$ that the system $A_R$ will operate until some moment $t$ (inclusive) depends on whether the system will operate at all moments $\tau(\tau=1,\ldots,t)$. The sequence of independent trials schema with operating probability $P_v(\tau)$ provides the reliability $R(t)$ defined as [1]:

$$R(t)=\prod_{\tau=1}^{t} P_v(\tau) \tag{2.2}$$

and further

$$1-\sum_{\tau=1}^{t}(1-P_v(\tau)) \leq R(t) \leq \exp(-\sum_{\tau=1}^{t}(1-P_v(\tau))) \tag{2.3}$$

Taking into consideration that $P_v(\tau) \to 1$ with $n(\tau)$ increase, there are possible different extreme values of $R(t)$ with $t$ increase. To refine this point, consider the ideal system $A_R$ with postulated features as follows [1]:

1. Operating capability. The system is capable to operate at any $\tau$ time moment. However if it fails at given time moment, then nothing can bring it into the operating state.
2. Unlimited extension. If the system is operating at given $\tau$ time moment, then at the next moment $\tau+1$ it may be enhanced by any number of elements. One time unit is a conditional one for a given system.
3. Physical restriction of reaction time. The system is got aware about its state only at the next $\tau+1$ time moment.
4. Math restrictions. The symmetric system with independent success and failure trial results at every given $\tau$ time moment is under consideration.

The limit of the reliability $R(t)$ of the symmetric system of $s-th$ degree with $n=const$ elements that are pair wise independent and uniformly distributed is defined by equation

$$R(t)=\exp(-\lambda \cdot t) , \tag{2.4}$$

where $\lambda$ is a system failure intensity measured in faults per element and is evaluated according to the systems' potential reliability law (see math proof in [1]) as follows:

$$-\ln(1-\exp(-k_L \cdot n)+O(\ln(n))) \leq \lambda \leq -\ln(1-\exp(-k_U \cdot n)) , \tag{2.5}$$

where

$$k_L = c \cdot \ln(\frac{c}{p_M}) + (1-c) \cdot \ln(\frac{1-c}{1-p_M}) ; \tag{2.6}$$

$$k_U = c \cdot \ln\left(\frac{c}{p_S}\right) + (1-c) \cdot \ln\left(\frac{1-c}{1-p_S}\right) \quad ; \qquad (2.7)$$

$$c = \frac{s}{n} \quad ;$$

$$p_M = \frac{p_L}{1 + p_L - p_U} \quad ;$$

$$p_M = p_L \leq p_v \leq p_U < 0.5 \quad ;$$

$$p_L = min(p_v), (v=1, 2, \ldots, n) \quad ;$$

$$p_U = max(p_v), (v=1, 2, \ldots, n) \quad ;$$

$$p_S = \frac{1}{n} \cdot \sum_{v=1}^{t} p_v \quad . \qquad (2.8)$$

The postulated features above fit a system approach to software development process including testing and debugging in particular. The systems' potential reliability law application seems to be a fruitful enough approach to software reliability quantification.

## 3. Enough Sigma Software Test Coverage Approach

Software reliability is understood as a probability of failure free software operation in defined environment for a specified period of time. In general a failure is a deviation of operation results from customer requirements. The deviation is defined by correspondence between algorithm's specification and its software implementation. Quality of algorithm's subject matter specification influences software reliability throughout the development process and lifecycle of software product. To achieve continuous improvement of software engineering process, reliability requirements must be defined in an integrated manner for prediction, evaluation, validation, verification, and certification at specification, coding, testing, maintenance, and correction cycles. Thus reliability monitoring needs to be implemented as an online automated process throughout the software lifecycle.

At the beginning we know nothing in general about algorithm to be implemented, but some ideas concerning input data and results. Consider algorithm's specification based on input data formal definition. Usefulness of a software reliability model depends on the definition method of input data to be tested for exhaustive fault detection. Input data may be considered as a set of requests (sites), so their total number and variety are sufficient for reliability evaluation. Sites are viewed as structural and(or) functional software elements including subject matter data, input variables, decisions, restrictions, memory structures and the like. All sites are potentially fault inherent and may cause a failure. Software input data set may be viewed as a site set. Software site is somewhat like a software path.

Let the site set structure defines total sites' number $n(\tau)$ at different $\tau$ time of lifecycle and occurrence probability $p_v(\tau)(v=1,2,\ldots,n(\tau))$ of $v-th$ site to be processed, that defines failure probability $1-p_v(\tau)$ while processing this site. Failure probability is greater for the sites that may occur more rarely, because it is more difficult in general to sensitize faults by such exotic sites while testing.

It is mostly improbable that all $n(\tau)$ sites will yield specified results because it is impossible to implement software without faults. But even if all $n(\tau)$ sites will yield specified results, this fact is undetectable because in this case we need all $n(\tau)$ sites to be tested. Practically it is impossible because of great values of number $n(\tau)$ for almost any software product. Not all sites are to be processed even throughout software lifecycle. To yield specified results at the required reliability level, in practice, it is enough for software product to have only $s(\tau)$ number of assured fault free sites. The number $s(\tau)$ of potentially faulty (sensitive) sites, sensitized throughout testing by time $\tau$, defines a software test coverage as $c(\tau)=s(\tau)/n(\tau), (0<c(\tau)<1)$ and is a known parameter of software reliability models.

It is natural in general to consider a software to be operating at a given $\tau$ time moment, if there exists at least some fixed, before stated, minimal part $c(\tau)$ of operating (fault free) sites of total sites' number $n(\tau)$.

The features (see Chapter 2) provide insight into the test process in general, including fault correction and regressive testing procedures, thus refining software lifecycle process according to these postulated features.

Any software site may sensitize fault(s) and is a potentially fault inherent one. Consider a software as a system comprised of $n(\tau)$ sites (elements). Operating set of this system consists of states, each including more than $s(\tau)$ operating sites. Consider the sites are pairwise independent and uniformly distributed over all possible sites' number $n(\tau)=const$. Software reliability $R(\tau)$ is evaluated by known equation (2.4) as

$$R(\tau)=\exp(-\lambda(\tau)) , \qquad (3.1)$$

where $\lambda(\tau)$ is a software failure intensity measured in faults per site and is evaluated (see formulas (2.5),…, (2.8)) as follows:

$$-\ln(1-\exp(-k_L(\tau)\cdot n(\tau)))+O(\ln(n(\tau)))\leq\lambda(\tau)\leq-\ln(1-\exp(-k_U(\tau)\cdot n(\tau))) \quad (3.2)$$

where

$$k_L(\tau)=c(\tau)\cdot\ln(\frac{c(\tau)}{p_M(\tau)})+(1-c(\tau))\cdot\ln(\frac{1-c(\tau)}{1-p_M(\tau)}) ; \qquad (3.3)$$

$$k_U(\tau)=c(\tau)\cdot\ln(\frac{c(\tau)}{p_S(\tau)})+(1-c(\tau))\cdot\ln(\frac{1-c(\tau)}{1-p_S(\tau)}) ; \qquad (3.4)$$

$$c(\tau)=\frac{s(\tau)}{n(\tau)} , \quad (0\leq c(\tau)\leq 1) ;$$

$$p_M(\tau)=\frac{p_L(\tau)}{1+p_L(\tau)-p_U(\tau)} ;$$

$$p_L(\tau)\leq p_v(\tau)\leq p_U(\tau)<0.5 ;$$

$$p_L(\tau) = min(p_v(\tau)) \quad , \quad (v = 1, 2, \ldots, n(\tau)) \quad ;$$

$$p_U(\tau) = max(p_v(\tau)) \quad , \quad (v = 1, 2, \ldots, n(\tau)) \quad ;$$

$$p_S(\tau) = \frac{1}{s(\tau)} \cdot \sum_{v=1}^{n(\tau)} p_v(\tau) \quad . \tag{3.5}$$

We consider $O(\ln n(\tau))$ to have some constant value that must be defined at large values of $n(\tau)$, thus minimum and maximum failure intensity according (3.2), (3.3) and (3.4) correspondingly are as follows:

$$\lambda_{min}(\tau) = -\ln(1 - \exp(-k_L(\tau) \cdot n(\tau)) + O(\ln(n(\tau)))) \tag{3.6}$$

$$\lambda_{max}(\tau) = -\ln(1 - \exp(-k_U(\tau) \cdot n(\tau))) \quad . \tag{3.7}$$

The Agile Software Reliability Monitoring Model (ASRMM) considers a software as a system, comprised of $n(\tau)$ sites. Operating set of this system consists of states, each including more than $s(\tau)$ operating sites. Software test process is defined in general according to the postulated features (Chapter 2), including fault correction and regressive testing, thus refining software lifecycle process as for achieving and supporting reliability requirements according to quantitative interdependences (3.1), (3.4), (3.7) of total sites' number $n(\tau)$, test coverage $c(\tau)$, mean site's occurrence probability $p_S(\tau)$, and failure intensity $\lambda(\tau)$ (see Figure 2). Time $0 \leq \tau \leq 1$ flow in the ASRMM is a math one and the $\tau-unit$ is defined as $\tau_s = \frac{1}{n(\tau)}$.

To define the initial mean occurrence probability $p_S(0)$, imagine a software site $r_v(v=1,\ldots,n(\tau))$ as a set of $W(\tau)$ pairwise independent parameters $x_i(i=1,\ldots,W(\tau))$. Each parameter may accept value(s) $x_{ij}(j=1,\ldots,J_i(\tau))$ with $s_{ij}(\tau)$ sensitive values' number of each. These parameters define semantically sufficient parameter's values that are sensitive as for yielding specified results, or, otherwise, values that may sensitize faults in software under test. These values are supposed to yield semantically typical results. Software sites' set structure may be viewed as a software sensitive sites' semantics matrix (SSSSM).

Semantics of the parameter values is defined by subject matter and the particularities of the algorithm's implementation. There may be defined a number $J_i(\tau)$ of semantic types having $s_{ij}(\tau)$, $(s_i(\tau) \geq s_{ij}(\tau) \geq 1)$ values' number of each type. So each parameter $x_i$ has a number $s_i(\tau) = \sum_{j=1}^{J_i(\tau)} s_{ij}(\tau)$ of values, thus a total sites' number is $n(\tau) = \prod_{i=1}^{W(\tau)} s_i(\tau)$ and the number of sensitive sites is $s_0(\tau) = \prod_{i=1}^{W(\tau)} J_i(\tau)$ for a given project. Then at the specification cycle $\tau = 0$, according to (3.5), we have $p_S(0) = \frac{1}{s_0(0)}$. Let's name $p_S(0)$ as an initial semantic mean of a software project under development. Values, denoted by $\tau$ time, vary throughout the development process due to refinements brought by customers, programmers, testers, developers, users, and the like. These refinements lead to the changes of total sites' number $n(\tau)$ and sensitive sites' number $s_0(\tau)$ because of detected faults, thus changing a value of semantic mean $p_S(\tau)$.

While testing, the test coverage value $c(\tau)$ increases so the difference $c(\tau) - p_S(0) < 0 \to 0$ and when
$$c(\tau) - p_S(0) = 0 , \qquad (3.8)$$
according to (3.4), we have $k_U(\tau) = 0$ that is a starting point of failure intensity (3.7) value decreasing and software reliability (3.1) growth (see Figure 2). When the sensitive sites' number $s_0(0)$ is not defined, e. g. in case of a black box testing, then according to (3.5) and (3.8) we have $\dfrac{s_0(0)}{n(0)} = \dfrac{1}{s_0(0)}$ and thus $s_0(0) = \sqrt{(n(0))}$.

Here is some general description of the software reliability monitoring algorithm based on the ASRMM.

Software site set structure is defined during specification as a SSSSM. Semantics of the site's parameter values is defined by software subject matter and algorithm's implementation particularities. The SSSSM may be viewed as a data base of the ASRMM engine (TESTERBOT) for generating tests, refining semantic mean and semantic shift of input data flow, thus providing monitoring of the software engineering process.

During software concept definition and input data specification at $\tau = 0$, before testing, we refine $s_0(0)$ sensitive sites selected from $n(0)$ total sites' number. Sites are ranged according to their occurrence probabilities $p_v(0), (v = 1, 2, \ldots, n(0))$ thus defining initial semantic mean $p_s(0)$ of input data flow.

Potential reliability metrics, evaluated according to (3.7), if compared with required ones $\lambda_{rq}(1)$, define the appropriate test coverage value to be achieved to meet the requirements. The difference $\lambda_{rq}(1) > \lambda(0)$ means that it is necessary to define the additional number of sensitive sites for testing to assure the achievement of required reliability. The additional sensitive sites are defined and may be selected by the TESTERBOT either in exhaustive or extreme (random or semantic extrapolation) mode of the test process.

Test process must assure the achievement of required test coverage value. While testing, fault(s) is (are) detected and corrected thus changing sensitive sites' number $s_0(\tau + \tau_s) := s_0(\tau) + 1$ for each fault(s) and in general changing the number of values $s_i(\tau)$ of the sites' parameters. The sensitive sites parameters' values are either predefined or randomly selected according semantic ranges. If total sites' number $n(\tau)$ changes then $\tau$ time unit $\tau_s$ is recalculated. If fault(s) is (are) not detected, the number of tested sites must be changed $s(\tau + \tau_s) := s(\tau) + 1$ for each tested site thus increasing test coverage $c(\tau)$.

Value of time is calculated as $\tau := \tau + \tau_s$ after each tested fault free site.

Changes are calculated for $n(\tau)$, $s(\tau)$, $s_0(\tau)$, $s_i(\tau)$, $J_i(\tau)$, $W(\tau)$ thus refining semantic shift $p_s(0) - p_s(\tau)$. If total sites' number $n(\tau)$ is changed, then the time unit is to be refined and the time value is to be recalculated only after each tested faulty site being corrected and retested.

Reliability metric $\lambda(\tau < 1)$, being continuously evaluated as (3.7) throughout testing process and compared with required one, provides a possibility of making decisions on testing process. The achieved reliability level, being continuously monitored, depicts changes either in subject matter requirements or software implementation particularities, thus providing possibilities for **optimization** throughout engineering process of software product development.

The ASRMM as a foundation of the Testerbot engineering yields **validated** and **verifiable** reliability evaluations, integrating subject matter data and quantitative software reliability metrics. Testerbot engineering provides continuous reliability evaluations throughout software engineering process, thus **accounting and tracing quantitative software reliability requirements** from customer to product.

Consider the customer's four sigma reliability requirements for the software project of the total number as a trillion ($10^{12}$) software sites of input data set's elements. At the beginning of the concept definition, the TESTERBOT preliminary evaluations yield the minimum needed software test coverage as 0.0000010032 or 1,003,200 tests to assure the achievement of required four sigma software reliability (0.00621 faults per site). To improve the reliability up to **six sigma** ($2.0 \cdot 10^{-9}$ faults per site), the extra needed **schedule-budget spending** on testing must be increased 1.003 times. Minimum failure intensity for the given $n(\tau)$ is $\lambda_{min}(\tau) = \frac{1}{n(\tau)}$, thus defining **Enough sigma** test coverage estimations to achieve maximum possible reliability for the given software project. To achieve **maximum** (Enough sigma) reliability, the engineering efforts must be 1.0042 fold. It seems that Six Sigma reliability may be not enough for large scaled, semantically sophisticated, and critical software projects.

During further software input data set structure refinement and specification at $\tau = 0$, before testing, the TESTERBOT refines initial number $s_0(0)$ of sensitive sites selected from the total $n(0)$ sites' number. Sites are ranged by the TESTERBOT according to their occurrence probabilities, thus refining initial semantic mean $p_s(0)$ of input data. So the TESTERBOT evaluations yield the refined minimum needed software test coverage as 0.0000010016 or 1,001,600 tests to assure the achievement of required four sigma software reliability. To improve the reliability up to six sigma, the extra needed schedule-budget spending on testing must be increased 1.00156 times. To achieve Enough sigma reliability, the engineering efforts must be 1.00211 fold.

The TESTERBOT application provides 0.37 % **schedule-budget savings** or 3720 tests' number decrease to achieve Enough sigma reliability requirements for the given project example.

The ASRMM results featuring the Testerbot **optimization** capabilities, give an idea of **schedule-budget-reliability estimations** of the needed test coverage changes due to refinements of total software sites' number. These refinements are contributed to the project by customer, developers, testers, and the like throughout the software engineering process.

Testerbot engineering assures the achievement of required software test coverage value thus defining the mostly accurate semantic mean compliance with semantic target. Any detected and corrected fault decreases semantic shift. While testing, fault(s) is (are) detected and corrected thus changing initial sensitive sites' number $s_0(0)$ and in general changing total sites' number $n(0)$, thus refining needed test coverage $c(\tau)$ to achieve required failure intensity $\lambda(\tau)$ for any sigma software reliability.

Figure 2 displays the interdependence of test coverage $c(\tau)$ and failure intensity $\lambda(\tau)$. The $\tau$ notation will be omitted for the example below so that e.g. $n(\tau) = n$.

Let the input data set is defined and numbered as integers $x = 1, 2, \ldots, 20$ comprising the total software sites' number $n = 20$ for a project example to execute, e. g. some function $y = f(x)$ under restrictions $5 \leq x < 10$ and $10 < x \leq 15$. The flowchart of this example (Figure 1) is comprised of four initial sensitive sites (branches) $y_1, y_2, y_3, y_4$ with corresponding occurrence probabilities $p_1 = p_2 = \frac{4}{20} = 0.2$, $p_3 = \frac{1}{20} = 0.05$, $p_4 = \frac{9}{20} = 0.45$. So initial sensitive sites' number $s_0(0) = 4$ yields according to (3.5) a semantic mean sites' occurrence probability $p_S = \frac{1}{4} = 0.25$.

Consider the minimum failure intensity for $n=20$ is $\lambda=\frac{1}{n}=0.05$, thus defining according to (3.2), (3.4) the Enough sigma test coverage estimation $c=0.55$ or that $\frac{0.55}{0.05}=11$ tests (Figure 2 in emerald) are needed to assure the achievement of required Enough sigma software reliability for the given software project example (Figure 1). Being considered as white boxes, the branches $y_1, y_2, y_3$ may be tested by sensitizing one site of each branch. So the black box $y_4=f(x)$ needs 11-3=8 tests to assure the Enough sigma reliability level of the project. Being considered as a white box, the branch $y_4$, e. g. $f(x)=\frac{1}{x-10}$, needs one test and to assure the required reliability $\lambda \leq 0.05$ for the given project the extrapolated test coverage (Figure 2 in gold) may be estimated as $c_e = \sum_{i=1}^{4}(p_i - \frac{1}{20})$.

The Testerbot software engineering depicts real software test process and may be viewed as an online TESTERBOT software development tool (PLM solution) or integrated development environment (IDE) to be applied for any test phase and strategy including extreme (agile) and exhaustive ones;

Each software product, released under the TESTERBOT tool, may be equipped with reliability e-certificate needed for acquisition, remedial and continual improvement processes, thus suggesting **the quantifiable improvement approach to system development standards**.

## 4. Content Relevance Quantification Approach

Any content sensitizes sense or results while processing by brain or computer. Content search aims to provide access to the results of content processing. These results are to be of enough quality to rely on, so the content search must be reliable enough to provide (detect) relevant content for processing.

Reliability of content search may be viewed as a probability of relevant content detection or coverage by applying a search query. Irrelevance may be considered as a deviation of content coverage results from customer requirements. The deviation may be defined as a compliance of the query and content specifications with search engine implementation. Quality of subject matter query specification influences the content search relevance.

In general we know nothing about content to be searched, but some ideas concerning queries and content searching results.

The **Content Relevance Quantification Model (CRQM)** considers any content as a set of $n$ queries with queries' variety number $s$ or sensitive queries' number $s \leq n$. Any content query $q_v (v=1,2,...,s)$ may sensitize sense and is a potentially sense inherent one. Input query, applied for sense detection (sensitization), may discover or recover content with some relevance to the search results.

Queries are viewed as structural and (or) functional content (document, collection, corpus) elements, including subject matter data. Let the query set structure defines sensitive queries' number $s$ and an occurrence probability $p_v, (v=1,2,...,s)$ of $v-th$ query, that defines a probability of sensitizing sense in response to this query.

It is mostly improbable, that an input query includes all $s$ sensitive queries specifying content under search, because of the great values of number $s$ for almost any content. To yield specified search results at the required relevance level, in practice, it is enough for a content to have only $s_v$ sensitive queries. The number $s_v$ of potentially sensitive queries, sensitized throughout content search, defines content's semantic coverage as $c_v = \frac{s_v}{n}$ and is a known parameter of search models.

It is natural to consider any content under search to be relevant enough, if there exist at least some fixed, before stated number $s_v$ of sensitive queries as a minimal part $c_v$ of the total content queries' number $n$.

Consider any content as a system comprised of $n$ queries, sensitive set of this system consists of states, each including more than $s_v$ sensitive queries. Consider queries are pairwise independent and uniformly distributed over all possible queries' number $n = const$. Content relevance $R$ or probability of sensitizing sense is quantified according to (2.4) by the equation $R = \exp(-\lambda)$, or else according to (2.5) as

$$R = 1 - \exp(-k_U \cdot n) , \qquad (4.1)$$

where $\lambda$ is a content's irrelevance intensity, defined by (2.5), …, (2.8) and measured as irrelevance (probability of not sensitizing sense) per content query.

Content query set structure is defined while indexing as a content query semantics matrix (CQSM). Semantics of the query's parameter values is defined by content subject matter and search engine's algorithm implementation particularities. The CQSM may be considered as a semantic quantifier's (SEQUANTIC tool) data base for queries' indexing (quantification), refining content's semantic mean $p_S$, semantic coverage $c_v$, and semantic shift $c_v - p_S$ thus providing content relevance quantification. Content search assures the achievement of required semantic coverage value.

Consider a content of total queries' number $n$ and sensitive queries' number $s$ with semantic mean $p_S = \frac{1}{s}$. The mostly relevant query for any content is the input query yielding $R_v = 1$, $\lambda_v = 0$. The mostly irrelevant (uncertain) search query $R_v \to 0$ in general is defined at $c_v = p_S = p_v$ and $\lambda_v \to \infty$, so that semantic coverage is equal to semantic mean and occurrence probability of the query $q_v$. The query **discovers** a content with higher relevance, the greater is the inequality $c_v < p_S$. The query **recovers** a content with higher relevance, the greater is the inequality $c_v > p_S$ (see Figure 3).

Consider as an example the content:

МУУ КУ КА РЕ КУ МУУ КУ КА РЕ КУ МУУ КУ КА РЕ КУ МУУ КУ КА РЕ КУ

of the unknown grammar and subject. This content may be structured as one-term total queries' (tokens') number $n = 20$ and the sensitive queries' number $s = 4$ with semantic mean $p_S = \frac{1}{s} = 0.25$. The occurrence probabilities $p_v, (v = 1, 2, …, 4)$ for sensitive tokens (queries) МУУ, КУ, КА, РЕ are consequently appear as $p_1 = \frac{4}{20} = 0.2$, $p_2 = \frac{8}{20} = 0.4$, $p_3 = \frac{4}{20} = 0.2$, $p_4 = \frac{4}{20} = 0.2$. Thus content's semantic coverage for these tokens are consequently equal to $c_1 = 0.2$, $c_2 = 0.4$, $c_3 = 0.2$, $c_4 = 0.2$ (see Figure 3). According to (4.1), (2.5),…, (2.8), the discovery relevance of this content for the query "МУУ" is $R_1 = 0.1306$ (see Figure 3 in emerald), and the recovery relevance, e. g. for the query "КУ" is $R_2 = 0.6611$. The query "КУ КА РЕ КУ" recovers this content as $c = c_2 + c_3 + c_4 = 0.4 + 0.2 + 0.2 = 0.8$ and is $R = 0.9999$ relevant to the content example (see Figure 3 in gold).

The presented content relevance quantification paradigm may be applied for the human grammar content after solving the known natural language ambiguities.

Content relevance $R$, being continuously evaluated throughout a search process and compared with required one, provides a possibility for making decisions on searching. The achieved relevance level depicts changes either in subject matter requirements (input queries) or search engine implementation particularities, thus providing possibilities for optimization throughout content search engineering process.

The **CRQM** improves latent semantic indexing, especially for unknown and (or) heterogenous collections, by increasing relevance, precision, and recall of content search, including the full text search. The **CRQM** may be used for data exploration and data integration tasks (due to its potential accuracy to quantify the content's semantics), to solve heterogeneity problems, and to provide varied levels of Querying services, that facilitates knowledge discovery at different levels of granularity.

The **CRQM** based SEQUANTIC tool as a content product lifecycle management solution may be viewed as a test bed or (and) a cradle for many existing or up coming content relevance models and search engine implementations.

Each content product, released under the SEQUANTIC tool, may be equipped with relevance e-certificate needed for acquisition, remedial and continual improvement processes, thus suggesting **the quantifiable improvement approach to the** content **development standards**.

The **CRQM** based content engineering yields **validated** and **verifiable** relevance estimations, integrating subject matter queries and quantitative relevance metrics of content. This engineering provides continuous relevance evaluations throughout content engineering process, thus **accounting and tracing quantitative** content **relevance requirements** from customer to the content product.

### 5. Conclusion. Towards Quantitative Software Reliability Grammar

Many research and development points may be initiated by the presented quantitative system engineering paradigm that provides any affordable quantitative accuracy in the systems' elements refinement suitable for the customer and the developer. One development point is a grammar of the language integrating software and content engineering throughout all development cycles.

Imagine a compiler parsing (browsing) through a source language program in search for sensitive sites to calculate their occurrence probabilities thus defining their semantic coverage to be applied for test coverage extrapolation in measuring the level of the achieved software project reliability. So the quantitative software reliability grammar of the programming language must markup the pairwise independent software subject matter parameters and sensitive values of these parameters to create test data base for quantitative monitoring of the development cycles. Such monitoring may be implemented as an enhancement of the integrated development environment for any programming language.

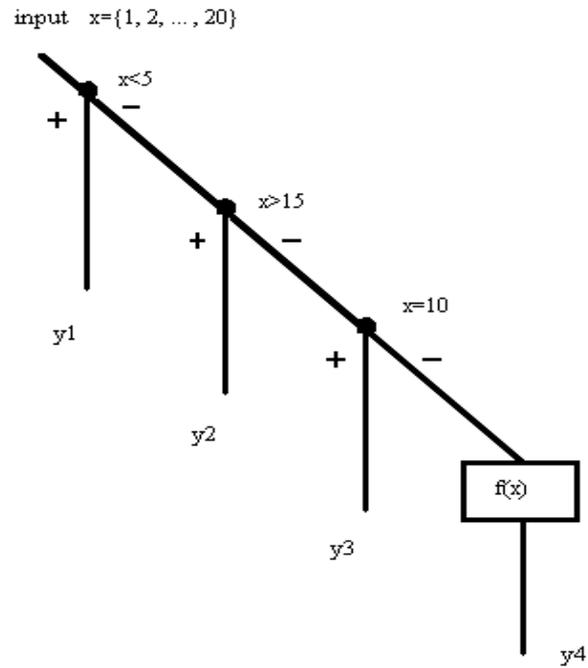

Figure 1. Flowchart of a software example project

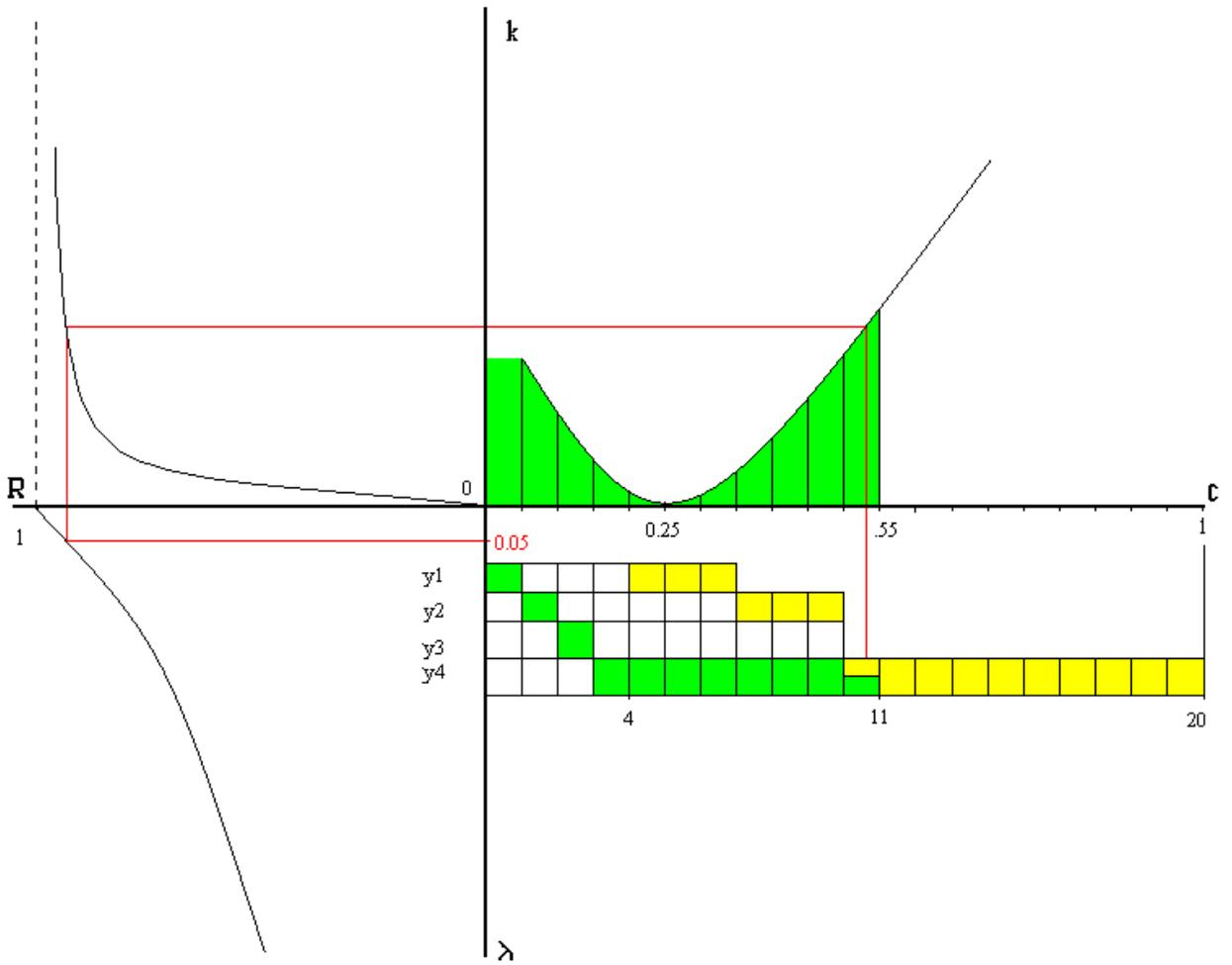

Figure 2. Quantitative interdependence of software test coverage $c$, software reliability $R$ and failure intensity $\lambda$.

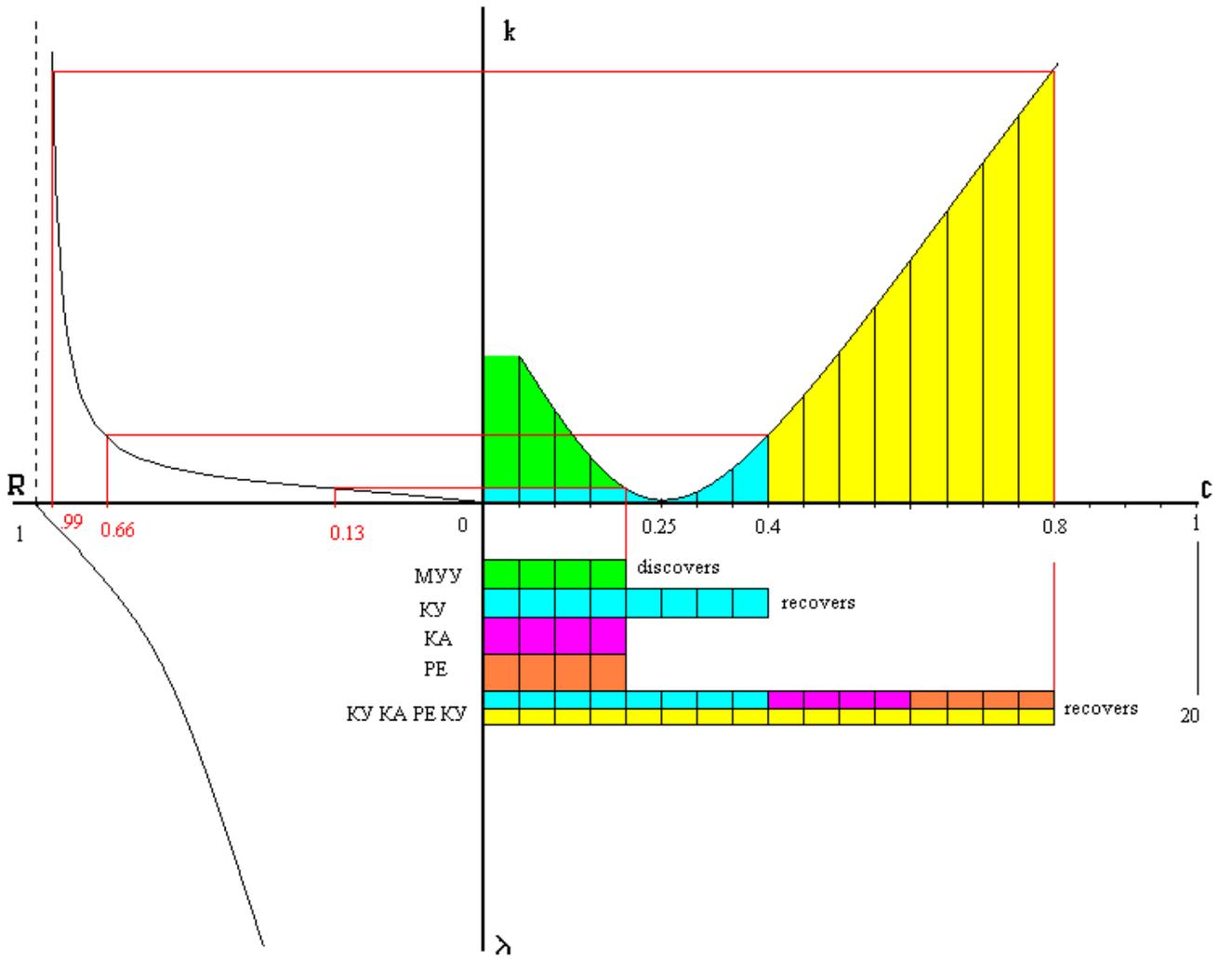

Figure 3. Quantitative interdependence of semantic coverage *c* and content relevance *R*